\documentstyle[preprint,aps]{revtex} 
\tighten
\begin{document}
\draft
\title{Hamilton-Jacobi Formulation of KS Entropy 
for Classical and Quantum Dynamics}
\author{M. Hossein Partovi \cite{byline}}
\address{Department of Physics and Astronomy, 
California State University, Sacramento, California 
95819-6041}
\date{\today}
\maketitle
\begin{abstract}

A Hamilton-Jacobi formulation of the Lyapunov 
spectrum and KS entropy is developed.  It is 
numerically efficient and reveals a close relation 
between the KS invariant and the classical action.  
This formulation is extended to the quantum domain 
using the Madelung-Bohm orbits associated with the 
Schroedinger equation.  The resulting quantum KS 
invariant for a given orbit equals the mean decay 
rate of the probability density along the orbit, 
while its ensemble average measures the mean growth 
rate of configuration-space information for the 
quantum system.  

\end{abstract}
\pacs{05.45.Mt, 03.65.Ca, 03.67.-a, 05.45.Pq}
\narrowtext
Chaos is a long-term instability exhibited by a 
dynamical system.  It is quantitatively measured by 
the Lyapunov spectrum of characteristic exponents, 
which represents the principal rates of orbit 
divergence in phase space, or alternatively by the 
Kolmogorov-Sinai (KS) invariant, which quantifies 
the rate of information production by the dynamical 
system \cite{1}.  The KS invariant is a fundamental 
quantity that underlies the information dynamics of 
a broad class of dynamical systems well beyond 
classical mechanics.  The results reported here 
show that it is intimately related to the classical 
action for Hamiltonian systems and to amplitude 
decay and 
configuration-space information for quantum 
systems.  While chaos is a ubiquitous feature of 
nonlinear classical dynamics, it is conspicuously 
absent in finite quantum systems as these are known 
to be dynamically stable \cite{2}.  Nevertheless, 
the chaotic nature of a given classical Hamiltonian 
produces certain characteristic features in the 
dynamical behavior of its quantized version.  These 
features have been studied extensively and are 
commonly referred to as ``quantum 
chaos''\cite{2.5,3}.  They include short-term 
instabilities and diffusive behavior versus 
dynamical localization, characteristic spectral 
statistics, wave function scarring, and other 
effects.  A quantitatively precise formulation of 
quantum dynamical instability smoothly straddling 
the quantum-classical transition, on the other 
hand, is at present lacking and is the subject of 
this work.   

This Letter presents a general approach to the 
information dynamics of the quantum-classical 
transition based on the Hamilton-Jacobi (HJ) 
formalism.  The formulation is exact and based on 
phase-space concepts, with the KS invariant playing 
a central role.  The extension to the quantum 
domain is accomplished by means of the orbits 
introduced by Madelung\cite{4} and later 
rediscovered by Bohm\cite{5}.  These quantum orbits 
are natural extensions of the classical phase-space 
flow to quantum mechanics, and provide the required 
bridge across the transition.  The present approach 
yields several new results, including (i) a 
powerful and numerically efficient method for 
calculating the KS invariant for classical 
Hamiltonian systems, (ii) the striking finding that 
the quantum KS invariant for a given Madelung-Bohm 
(MB) orbit is equal to the mean decay rate of the 
probability density along the orbit, (iii) the 
result that the quantum KS invariant averaged over 
the ensemble of MB orbits equals the mean growth 
rate of configuration-space information, and (iv) a 
general and rigorous argument for the conjecture 
that the standard quantum-classical correspondence 
(or the classical limit) necessarily breaks down 
for classically chaotic Hamiltonians.  Several 
other results, including those of numerical 
simulations for classical and quantum models, are 
presented and discussed.  

{\bf Canonical formulation of chaos}.  Our 
objective here is a symplectically reduced 
formulation of the Lyapunov spectrum for a 
Hamiltonian system in terms of the classical 
action; see Ref. \cite{6}, hereafter referred to as 
``Paper I,'' for definitions and notation.  
Consider a classical system of $N$ degrees of 
freedom described by the canonical variables $\{ 
{q}_{i},{p}_{i} \}$, $i=1, \ldots,N$, the 
Hamiltonian function $H({\bf q},{\bf p},t)$, and 
Hamilton's principal function $S({\bf q},t,{{\bf 
p}}_{0})$, with ${{\bf p}}_{0}$ denoting the 
initial momenta\cite{7}.   Hamilton's equations can 
be compactly stated in matrix form as 
${\dot{\bbox{\xi}}}={\cal J} {\bbox{\nabla}}_{\xi} 
H({\bbox{\xi}},t)$, where ${\bbox{\xi}}$ stands for 
the $2N$-dimensional phase-space vector $({\bf 
q},{\bf p})$.  Here $\cal J$ is a real, 
antisymmetric matrix of order $2N$ with a $2 
\otimes N$ block form $({\it 0}_{N}, {I}_{N},-
{I}_{N},{\it 0}_{N})$, which is simply a listing of 
its blocks in the order $(11,12,21,22)$.  
The tangent dynamics of the system is described by 
the $2N \times 2N$, nonsingular matrix ${\cal 
T}_{\mu \nu}(t,{\bbox{\xi}}_{0})\stackrel{\rm 
def}{=}\partial {{\xi}}_{\mu} (t,{\bbox{\xi}}_{0}) 
/ \partial {{\xi}}_{0 \nu}$, the {\it sensitivity} 
matrix, where ${\bbox{\xi}} (t,{\bbox{\xi}}_{0})$ 
is the trajectory that starts from 
${\bbox{\xi}}_{0}$ at $t=0$.  

Our first task is to express ${\cal T}$ in terms of 
Hamilton's principal function $S$.  Using the 
properties of $S$, we find the desired result in 
the block form 
\begin{eqnarray}
{\cal T}=(&&{S_{{\bf p}_{0}{\bf q}}}^{-1},-{S_{{\bf 
p}_{0}{\bf q}}}^{-1}S_{{\bf p}_{0}{\bf 
p}_{0}},\nonumber\\
&&S_{{\bf q}{\bf q}}{S_{{\bf p}_{0}{\bf q}}}^{-
1},{\tilde{S}}_{{\bf p}_{0}{\bf q}}- {S}_{{\bf 
q}{\bf q}}{S_{{\bf p}_{0}{\bf q}}}^{-1}S_{{\bf 
p}_{0}{\bf p}_{0}}),  \label{0.1}
\end{eqnarray}
where ${(S_{{\bf p}_{0}{\bf q}})}_{ij}\stackrel{\rm 
def}{=}{\partial}^{2}S/ {\partial}_{{q}_{j}} 
{\partial}_{{p}_{0i}}$\cite{7.5}.  Note that ${\cal 
T}_{21}{{\cal T}_{11}}^{-1}$ is a symmetric matrix 
(it equals $S_{{\bf q}{\bf q}}$), a fact that is a 
direct consequence of the symplectic symmetry of 
the underlying Hamiltonian, which symmetry implies 
the condition ${\cal T}{\cal J}{\tilde{\cal 
T}}={\cal J}$\cite{6}.  This allows us to transform 
the sensitivity matrix ${\cal T}$ to an 
upper-triangular block form $\Gamma$ according to 
$\Gamma\stackrel{\rm def}{=}\Omega (\Theta) {\cal 
T}$, where $\Omega(\Theta) \stackrel{\rm 
def}{=}(\cos \Theta, -\sin  \Theta,\sin \Theta, 
\cos \Theta)$ and $\tan (\Theta) \stackrel{\rm 
def}{=}-S_{{\bf q}{\bf q}}$\cite{7.6}.  Note that 
$\Theta$ is a real, symmetric matrix of order $N$, 
while $\Omega$ is orthogonal and symplectic.  We 
shall refer to $\Theta$ as the {\it symplectic 
phase} matrix for reasons that will become clear in 
the following.

The upper-triangular form $({\Gamma}_{11}, 
{\Gamma}_{12}, {\it 0}_{N}, {\Gamma}_{22})$ of 
$\Gamma$, on the other hand, guarantees that 
${\Gamma}_{11}^{-1}= {\tilde{\Gamma}}_{22}$, and 
that the upper (lower) half of the Lyapunov 
spectrum is obtained from the singular values of 
${\Gamma}_{11}$ (${\Gamma}_{22}$) as explained in 
Paper I.  In particular, the KS entropy is given 
by\cite{7.7} 
\begin{equation}
k={\lim \,}_{t \rightarrow \infty }  \log[ \det 
({\Gamma}_{11})]/t. \label{0.2}
\end{equation}

The evolution equation for $\cal T$, Eq.~(\ref{1}) 
of Paper I, leads to the set of matrix equations
\begin{equation}
{\dot {\Gamma}}_{11} {\Gamma}_{11}^{-
1}={L}^{A}+{\cal K}^{\Omega}_{21}$, \, ${\dot 
{\Omega}}={\cal L} \Omega$, \, ${L}^{S}+ {\cal 
K}^{\Omega}_{11} =0, \label{0.3}
\end{equation}
which can be used to calculate the Lyapunov 
spectrum efficiently.  Here ${\cal L}\stackrel{\rm 
def}{=}({L}^{A},{L}^{S},-{L}^{S},{L}^{A})$, $ {\cal 
K} (t,{\bbox{\xi}}_{0})\stackrel{\rm 
def}{=}{\bbox{\nabla}}_{\xi} {\bbox{\nabla}}_{\xi} 
H[{\bbox{\xi}}(t,{\bbox{\xi}}_{0}),t]$, and ${\cal 
K}^{\Omega} \stackrel{\rm def}{=}{\Omega}{\cal 
K}{\tilde{\Omega}}$, where ${L}^{S}$ and ${L}^{A}$ 
are symmetric and antisymmetric matrices 
respectively.  In particular, we note that the time 
average of the trace of the first equation in the 
set yields the KS invariant, i.e., $k=<{\rm 
tr}({\dot {\Gamma}}_{11} {\Gamma}_{11}^{-1})>$, 
where $<f> \stackrel{\rm def}{=} {\lim \,}_{t 
\rightarrow \infty } {t}^{-1} {\int}_{0}^{t} 
f(\bar{t}) d\bar{t} $.  Using the above set of 
equations, we find
\begin{equation}
{\dot {\sigma}}+{\cal K}_{11}+{\cal 
K}_{12}{\sigma}+{\sigma}{\cal K}_{21}+{\sigma}{\cal 
K}_{22}{\sigma}=0,
\label{1}
\end{equation}
\begin{equation}
k=<{\rm tr} [ {{\cal K}_{11}-{\cal K}_{22} \over 2} 
\sin 2 \Theta + {{\cal K}_{12}+{\cal K}_{21} \over 
2} \cos 2 \Theta ]>,
\label{2}
\end{equation}
where ${\sigma}(t)$ is the restriction of 
${S}_{{\bf q}{\bf q}}=-\tan (\Theta)$ to a specific 
orbit and ${\sigma}(0)=0$.  Equations~(\ref{1}) and 
(\ref{2}), together with the equations of motion 
for the orbit, constitute a closed set from which 
the KS invariant can be calculated for any 
Hamiltonian system.  The main computational burden 
is in solving the $N(N+1)/2$ differential equations 
in Eq.~(\ref{1}).  As with the results of Paper I, 
the above set of equations is free of exponentially 
growing quantities and does not require any 
orthonormality maintenance.  

To illustrate the power of these equations, let us 
consider the standard form $H={\bf p}^{2}/2 + 
V({\bf q},t)$, where $\bf q$ and $\bf p$ are 
$N$-dimensional vectors.  We then  find ${\cal 
K}_{11}={\bbox{\nabla}}_{q} {\bbox{\nabla}}_{ q} 
V({\bf q},t)$, ${\cal K}_{12}={\cal K}_{21}={\it 
0}_{N}$, ${\cal K}_{22}={I}_{N}$, and 
\begin{equation}
{\dot {\sigma}}+{\sigma}^{2}+{\cal K}_{11}=0,
\label{3}
\end{equation}
\begin{equation}
k={1 \over 2}<{\rm tr} ({\cal K}_{11}-1)\sin 2 
\Theta>.
\label{4}
\end{equation}
Moreover, using $\sigma=-\tan (\Theta)$ and 
Eq.~(\ref{3}), we can rewrite Eq.~(\ref{4}) in the 
remarkably simple form  
\begin{equation}
k={<{\rm tr}\sigma >}_{p.v.},
\label{5}
\end{equation}
where ``p.v.'' stipulates a principal-value 
evaluation\cite{8}.  Since ${\rm tr}(\sigma) $ 
equals ${\nabla}_{q}^{2}S $ along the orbit, 
Eq.~(\ref{5}) simply  states that {\it the KS 
invariant equals the time-average of the Laplacian 
of the action along the orbit}.

We will consider two examples here.  The first is 
defined by $N=3$ and $V({\bf q})=-{1 \over 2} 
[{({q}_{1}-{q}_{2})}^{2} +{({q}_{2}-
{q}_{3})}^{2}+{({q}_{3}-{q}_{1})}^{2}]+ 
({q}_{1}^{4}+2{q}_{2}^{4}+3{q}_{3}^{4})$.  
Equations\ (\ref{3}) and (\ref{4}) together with 
Hamilton's equations, $13$ in all, were integrated 
for a duration of $2.2 \times {10}^{7}$ time units 
with the initial value of $H$ set at unity\cite{9}.  
The result was $k=0.6126$, with the fluctuations 
never exceeding $0.0001$ as the simulation was 
extended by one order of magnitude to the endpoint.  
In fact a $1 \%$ result was already achieved at 
about ${10}^{4}$ time units.  

For a second example, we will derive explicit 
formulas for an $N$-dimensional kicked system 
(which is otherwise free) with $V({\bf q},t)=f({\bf 
q}){\sum}_{n=1}^{\infty} \delta (t-nT)$.  Let 
$({{\bf q}}_{n}, {{\bf p}}_{n})$ be the coordinates 
just after the ${n}^{th}$ kick.  Then $({{\bf 
q}}_{n+1}, {{\bf p}}_{n+1})= ({{\bf q}}_{n}, {{\bf 
p}}_{n})+[{{\bf p}}_{n}T,-{\nabla} f({{\bf 
q}}_{n+1})]$.  From Eq.~(\ref{3}), on the other 
hand, we find ${\dot {\sigma}}+{\sigma}^{2}=0$ for 
$t \neq nT$ and ${\Delta \sigma}_{n}=-{\nabla 
\nabla } f({{\bf q}}_{n})$ across the ${n}^{th}$ 
kick.  The corresponding solution is easily found 
to be ${\sigma}(t)={(t-nT+{\sigma}_{n}^{-1})}^{-
1},\ nT < t < (n+1)T$.  Thus the iteration map for 
$\sigma$ is \begin{equation}
{\sigma}_{n+1}={({\sigma}_{n}^{-1}+T)}^{-1}-
{\nabla} {\nabla} f({{\bf q}}_{n+1}), \label{5.9}
\end{equation}
where ${\sigma}(0)=0$.  We now use Eqs.~(\ref{5}) 
and (\ref{5.9}) to find
\begin{equation}
k={\lim \,}_{N \rightarrow \infty }{(NT)}^{-
1}{\sum}_{n=0}^{N-1} \ln |\det (1+{\sigma}_{n}T)|,
\label{6}
\end{equation}
which, together with the iteration maps given 
above, provides an efficient algorithm for 
calculating the KS invariant for kicked systems.  
As an illustrative example, we chose $ f({\bf q})=-
{1 \over 2} [{({q}_{1}-{q}_{2})}^{2} +{({q}_{2}-
{q}_{3})}^{2}+{({q}_{3}-{q}_{1})}^{2}]+ 
({q}_{1}^{4}+{q}_{2}^{4}+{q}_{3}^{4})$, with $T=1.0 
\times {10}^{-10}$.  The value $k=1.5 \times 
{10}^{5}$ was obtained in about ${10}^{7}$ 
iterations\cite{10}.  
  
Note, incidentally, that in case an eigenvalue of 
${\sigma}_{n}^{-1}$ is negative and less than $T$ 
in magnitude, $\sigma$ will encounter a simple pole 
in the ${n+1}^{st}$ cycle [other types of 
singularities are ruled out by Eq.~(\ref{3})].  
This behavior is generic for chaotic Hamiltonians 
(and common for all Hamiltonians), since otherwise 
a bounded $\sigma$ would, by virtue of 
Eq.~(\ref{5}), lead to $k=0$.  Correspondingly, the 
symplectic phase angles (eigenvalues of $\Theta$) 
go through $(m-{1 \over 2})\pi$ as $t$ goes through 
the ${m}^{th}$ singularity of that eigenvalue [we 
take $\Theta(0)=0$].  These poles occur when ${\cal 
T}_{11}$ becomes singular (as a matrix), and are 
the analogues of the so-called {\it conjugate} 
points which occur when ${\cal T}_{12}$ becomes 
singular\cite{2.5}.  Further insight is gained by 
considering these angles for quadratic potentials 
for which ${\cal K}_{11}$ is a constant, 
positive-definite matrix (and of course $k=0$).  
The solution to Eq.~(\ref{3}) is found to be 
$\sigma = -{\cal K}_{11}^{1 \over 2}\tan ( {\cal 
K}_{11}^{1 \over 2} t)$.  Thus the symplectic 
phases are given by 
${\theta}_{i}=\arctan[{\omega}_{i} \tan 
({\omega}_{i}t)]$, where $\{ {\omega}_{i} \}$ are 
the characteristic frequencies of the system.  
Clearly, the $m$th singularity associated with 
${\theta}_{i}$ occurs for ${\omega}_{i}{t}=(m-{1 
\over 2})\pi$ as expected.  The conjugate points 
occur midway between the latter at ${\omega}_{i}t=m 
\pi$.   

{\bf Extension to Quantum Mechanics}.  The MB 
formalism \cite{4,5} associates a phase space flow 
with a quantum system by setting\cite{11}
\begin{equation}
 \psi({\bf x},t) \stackrel{\rm def}{=} 
\exp[i{\sf{S}}({\bf x},t)/ \hbar + {\sf{R}}({\bf 
x},t)], 
\label{6.1}
\end{equation}
and treating ${\sf{S}}$ as Hamilton's principal 
function for an ensemble of orbits with 
configuration-space density ${\exp}(2{\sf{R}})$.  
These orbits are defined according to 
\begin{equation}
{\dot {\sf q}}(t,{\sf q}_{0},{\sf p}_{0})={{\sf 
p}}(t,{\sf q}_{0},{\sf 
p}_{0})={\nabla}{\sf{S}}[{\sf q}(t,{\sf q}_{0},{\sf 
p}_{0}),t], 
\label{6.2}
\end{equation}
where ${\sf q}_{0}{=} {\sf q}(0,{\sf q}_{0},{\sf 
p}_{0})$.  Here $\psi({\bf x},t)$ is a solution of 
the Schroedinger equation with the Hamiltonian 
${\bf p}^{2}/2 + V({\bf x},t)$.  It is readily 
verified that the expectation value of any 
observable in the state $\psi({\bf x},t)$ is given 
by its average over the ensemble of orbits defined 
above.  In particular, the ensemble average of 
Eq.~(\ref{6.2}) will lead to Ehrenfest's equations.  

The correspondence thus established allows us to 
define the quantum KS invariant for a given orbit 
as 
\begin{equation}
{\sf k}\stackrel{\rm def}{=}{<{\nabla}^{2} {\sf 
S}>}_{p.v.}, 
\label{6.25}
\end{equation} 
in strict analogy to Eq.~(\ref{5}).  It is worth 
emphasizing here that the averaging process in 
Eq.~(\ref{6.25}) is with respect to the time along 
the MB orbit to which $\sf S$ is restricted.     

Intuitively, one would expect that orbits 
neighboring a hypothetical chaotic orbit in the 
ensemble diverge from it on the average, thus 
causing the orbit density along the chaotic orbit 
to decrease with a mean rate related to $\sf k$.  
Remarkably, this expectation is fully realized by 
the Schroedinger equation.  To see this, consider 
the equation of motion for ${\sf{R}}({\bf x},t)$ as 
inherited from the Schroedinger equation:  
${\partial}{\sf{R}} / {\partial} t + 
{\nabla}{\sf{R}}{\cdot} {\nabla}{\sf{S}} = -{1 
\over 2}{\nabla}^{2} {\sf S}$.  An inspection of 
this equation shows that its characteristic curves 
are the MB orbits, so that it takes the following 
form along those orbits: 
\begin{equation}
d{\sf{R}}({\sf q}(t,{\sf q}_{0},{\sf p}_{0}),t) / 
dt = -{1 \over 2}{\nabla}^{2} {\sf{S}}[{\sf 
q}(t,{\sf q}_{0},{\sf p}_{0}),t].
\label{6.3}
\end{equation}
Using this in Eq.~(\ref{6.25}), we find ${\sf 
k}{=}-2{<{d {\sf{R}}}/ dt>}_{p.v.}$, or 
equivalently
\begin{equation}
{\sf k}{=}-{<d \ln  {|\psi[{\sf q}(t,{\sf 
q}_{0},{\sf p}_{0}),t]|}^{2} /dt >}_{p.v.}. 
\label{6.4}
\end{equation}
This exact result states that {\it the quantum KS 
invariant for a given orbit is the mean decay rate 
of the probability density along the orbit}.  We 
mention in passing that if $\psi$ in 
Eq.~(\ref{6.4}) is replaced with its classical 
limit ${\psi}_{s.c.}={[\det  (S_{{\bf p}_{0}{\bf 
q}})]}^{1 \over 2} \exp[i{{S}}({\bf x},t)/ \hbar]$, 
the result is the classical KS invariant 
$k$\cite{7.5}. 

Let us consider the implications of Eq.~(\ref{6.4}) 
for classically chaotic systems, for which $k \neq 
0$ while $\sf k =0$.  For a chaotic orbit in 
classical dynamics, there is no integral of motion 
in phase-space except possibly energy.  For the 
ensemble of MB orbits, on the other hand, the 
defining equations (\ref{6.2}) provide $N$ such 
integrals (upon eliminating ${\dot {\sf q}}$) if 
${\nabla}{\sf{S}}$ is a long-term predictable, or 
{\it computable}, function.  But ${\nabla}{\sf{S}}$ 
is computable, since the Schroedinger equation for 
finite-dimensional, bounded systems is known to be 
free of sensitivity to initial conditions\cite{2}.  
The inevitable conclusion is that while 
${\nabla}{{S}}$ (classical) is not computable for 
chaotic systems, ${\nabla}{\sf{S}}$ (quantum 
mechanical) is, regardless of how small ${\hbar}$ 
may be.  Clearly, the statement that 
${\nabla}{\sf{S}}$ reduces to ${\nabla}{{S}}$ as 
${\hbar} \rightarrow 0$, often referred to as the 
{\it classical limit}, cannot hold for chaotic 
Hamiltonians.  Recalling that regular Hamiltonians 
are very special while chaotic ones are by far the 
more typical class, we recognize the very limited 
validity of the classical limit. 

The above considerations confirm the view that the 
absence of sensitivity to initial conditions in the 
quantum dynamics of classically chaotic 
Hamiltonians originates in the breakdown of the 
classical limit itself for large times.  Stated 
otherwise, the two limiting procedures ${\hbar} 
\rightarrow 0$ and $t \rightarrow \infty$ do not 
commute for classically chaotic 
Hamiltonians\cite{2,3,12}.  How is this related to 
the issue of computability raised above?  Long-time 
unpredictability is precisely the condition that 
characterizes ${\nabla}{{S}}$ as uncomputable, in 
contrast to ${\nabla}{\sf{S}}$ which is computable 
by way of the Schroedinger equation.  Of course, 
for sufficiently small times the two can be 
arbitrarily close.  That a quantum system can 
temporarily exhibit dynamic instability was 
recognized early in a numerical study of the 
quantum kicked rotor model\cite{13}, and has since 
been investigated extensively\cite{3}. 

Indeed it is useful to restate the general 
arguments above in more explicit terms for this 
widely studied model.  Classically, the kicked 
rotor is a driven, one-dimensional system with 
chaotic orbits for appropriate values of the 
driving force.  The MB orbits for this system, on 
the other hand, are governed by (\ref{6.2}), which 
amounts to a single, non-autonomous first-order 
differential equation.  As such, this system cannot 
possibly be chaotic\cite{14}, implying that in the 
course of time the quantum orbits will deviate 
arbitrarily far from the classical ones.  Therefore 
the classical limit must of necessity fail for this 
system.  Correspondingly, $\sf k$ as measured 
according to Eq.~(\ref{6.25}) must vanish.  

In a numerical simulation of the quantum kicked 
rotor for the purpose of verifying this prediction, 
we integrated Eq.~(\ref{6.2}) together with the 
Schroedinger equation, and measured $\sf k$ along 
the resulting MB orbit using Eq.~(\ref{6.25}).  The 
result was zero\cite{15}, thus verifying the 
predicted regularity of the orbit.  An alternative 
calculation using a pair of nearby orbits 
reconfirmed the regularity of these orbits.  This 
raises the question of whether $\sf k$ can ever be 
positive for this model.  To answer this question, 
we explored the behavior of the hybrid quantity 
${k}^{hyb}\stackrel{\rm def}{=}{<{\nabla}^{2} {\sf 
S}[{\bf q}(t,{\bf q}_{0},{\bf p}_{0}),t]  
>}_{p.v.}$. Observe that here we are considering 
the {\it quantum} objects $\psi$ and ${\sf S}$ as 
evaluated along the {\it classical} orbit ${\bf 
q}(t,{\bf q}_{0},{\bf p}_{0})$ (in place of MB 
orbits).  Note that while ${k}^{hyb}$ is not 
required to vanish by the general arguments above, 
there is no a priori reason to the contrary either, 
despite the chaotic nature of the classical orbit.  
Numerical results, however, showed nonzero values 
of ${k}^{hyb}$ for a variety of conditions, in 
striking contrast to the behavior of $\sf k$.  This 
comparison clearly identifies the difference 
between the chaotic classical orbits and their 
quantum counterparts as the origin of the dynamical 
stability of the quantum kicked rotor.  It also 
indicates that the mean decay rate of $\psi[{\bf 
q}(t,{\bf q}_{0},{\bf p}_{0}),t]$, the restriction 
of the wavefunction to the classical orbit, is a 
discriminating footprint of classical chaos in 
quantum dynamics.  A systematic study of this 
intriguing phenomenon must await a better 
analytical understanding of the hybrid object 
$\psi[{\bf q}(t,{\bf q}_{0},{\bf p}_{0}),t]$.  

Thus far we have considered $\sf k$ for single MB 
orbits.  What is the MB ensemble average of $\sf 
k$, denoted by $\bar{\sf k}$?  Recalling that the 
MB ensemble average is the same as the quantum 
mechanical expectation value, we find from 
Eq.~(\ref{6.4}) or (\ref{6.25})
\begin{equation}
\bar{\sf k}={\lim \,}_{t \rightarrow \infty }  
{1 \over t} \int d{\sf q} {|\psi({\sf q},t)|}^{2}
\ln [{|\psi ({\sf q},t)|}^{-2}]. \label{9}
\end{equation} 
The integral occurring in (\ref{9}) is recognized 
as the information associated with the probability 
density ${|\psi({\sf q},t)|}^{2}$.  It is actually 
the entropy associated with an idealized position 
measurement, as shown in Ref. \cite{16}.   Thus we 
have arrived at the result that {\it the ensemble 
average of the quantum KS invariant equals the mean 
growth rate of the position measurement entropy}.  
Precisely this result was proposed in Ref. 
\cite{17} on the basis of more intuitive 
considerations.  The fact that the original 
information-theoretical significance of the KS 
entropy has naturally re-emerged in the quantum 
domain may be viewed as an affirmation of the 
present approach.

As is well known\cite{13,3}, short-term 
measurements of $\sf k$ typically show positive 
values.  In light of Eqs.~(\ref{6.4}) and 
(\ref{9}), this suggests a tendency of the 
probability density to decrease along certain MB 
orbits thereby developing information, i.e., 
structure and contrast, in configuration space.  
Presumably, points of low concentration would 
represent the interiors of regions of instability 
corresponding to classical chaos, with the 
boundaries as points of high concentration.  This 
scenario is reminiscent of the scarring phenomena 
discovered by Heller\cite{18} and merits further 
exploration.

This work was supported in part by a research grant 
from California State University, Sacramento.


\begin{references}
\bibitem{7.7} While in general $k$ is associated 
with some phase-space region and defined 
independently of the Lyapunov spectrum, we are 
relying on Pesin's result and the ergodicity of the 
chaotic orbit in writing Eq.~(\ref{0.2}); see Paper 
I.
\bibitem{8} In deriving Eq.(~\ref{5}), we have 
dropped the quantity ${\rm tr}< d 
\ln(1+{\sigma}^{2})/dt>$ which vanishes if $\sigma$ 
is bounded everywhere.  However, Eq.~(\ref{3}) 
implies the simple pole behavior $\sigma \sim {(t-
{t}_{s})}^{-1}$ near any singularity $t={t}_{s}$.  
Since the principal-value contribution to 
Eq.~(\ref{4}) from a simple pole vanishes, the 
result follows.
\bibitem{9} The simulation was performed on an HP 
9000 computer using the Bulirsch-Stoer method for a 
duration of $2.2 \times {10}^{7}$ units in $1.5 
\times {10}^{8}$ steps, with the local accuracy set 
at ${10}^{-10}$.   The starting point was 
${\bbox{\xi}}_{0}=(-0.075,0.1,-0.2,-
0.25,0.12,{\xi}_{8})$, where ${\xi}_{8} \geq 0 $ 
was determined from the energy condition.   
Stoermer's scheme was used for Hamilton's 
equations, resulting in an energy drift of less 
than $1.8 \times {10}^{-8}$ by the end of the 
simulation.
\bibitem{10} The starting point for this simulation 
was ${\bbox{\xi}}_{0}=(1,0,0,0,0,0)$.  The 
fluctuations remained at the $1$-$2$\% level as the 
simulation was extended to ${10}^{8}$ iterations.
\bibitem{14} Chaos is not possible with less than 
three autonomous, or two non-autonomous, 
first-order differential equations.
\end{references}
\end{document}